\documentclass{elsart5p}
\usepackage{color,setspace,times}
\usepackage{amssymb,amsmath,graphicx,color}
\usepackage[square,sort&compress,comma]{natbib}
\usepackage{lineno}

\begin{document}

% Chuang Liu, Xiu-Lei Jiang, Tao Liu, Ling Zhao, Wei-Xing Zhou, Wei-Kang Yuan (ECUST) %

\begin{frontmatter}

\title{Multifractal analysis of the fracture surfaces of foamed polypropylene/polyethylene blends}
\author[SB,RCSE]{Chuang Liu},
\author[SKLCE]{Xiu-Lei Jiang},
\author[SKLCE]{Tao Liu},
\ead{liutao@ecust.edu.cn}%
\author[SKLCE]{Ling Zhao},
\author[SB,RCSE,SS,RCE]{Wei-Xing Zhou\corauthref{cor}},
\corauth[cor]{Corresponding author. Address: 130 Meilong Road, P.O.
Box 114, East China University of Science and Technology, Shanghai
200237, China, Phone: +86 21 64253634, Fax: +86 21 64253152.}
\ead{wxzhou@ecust.edu.cn}%
\author[SKLCE]{Wei-Kang Yuan}

\address[SB]{School of Business, East China University of Science and Technology, Shanghai 200237, China}
\address[RCSE]{Research Center of Systems Engineering, East China University of Science and Technology, Shanghai 200237, China}
\address[SKLCE]{State Key Laboratory of Chemical Engineering, East China University of Science and Technology, Shanghai 200237, China}
\address[SS]{School of Science, East China University of Science and Technology, Shanghai 200237, China}
\address[RCE]{Research Center for Econophysics, East China University of Science and Technology, Shanghai 200237, China}

\begin{abstract}
The two-dimensional multifractal detrended fluctuation analysis is
applied to reveal the multifractal properties of the fracture
surfaces of foamed polypropylene/polyethylene blends at different
temperatures. Nice power-law scaling relationship between the
detrended fluctuation function $F_{q}$ and the scale $s$ is observed
for different orders $q$ and the scaling exponent $h(q)$ is found to
be a nonlinear function of $q$, confirming the presence of
multifractality in the fracture surfaces. The multifractal spectra
$f(\alpha)$ are obtained numerically through Legendre transform. The
shape of the multifractal spectrum of singularities can be well
captured by the width of spectrum $\Delta\alpha$ and the difference
of dimension $\Delta f$. With the increase of the PE content, the
fracture surface becomes more irregular and complex, as is
manifested by the facts that $\Delta\alpha$ increases and $\Delta f$
decreases from positive to negative. A qualitative interpretation is
provided based on the foaming process.
\end{abstract}


\end{frontmatter}

\section{Introduction}

The plastic foam industry is fast growing and the plastic foams have
drawn a great deal of interest in recent decades
\cite{Lee-Zeng-Cao-Han-Shen-Xu-2005-CoST}. The cell sizes of the
plastic foams strongly influence the application of the materials
\cite{Sun-Mark-Tan-Venkatasubramanian-2005-P}. The microcellular
polymer, with cell sizes hundreds of times smaller than those of
conventional plastic foams, can offer some unique properties that
conventional foams do not possess, such as higher impact strength,
higher toughness, higher stiffness-to-weight ratio, higher fatigue
life, higher thermal stability, lower dielectric constant, and lower
thermal conductivity
\cite{Nalawade-Picchioni-Janssen-2006-PPS,Tomasko-Li-Liu-Han-Wingert-Lee-2003-IECR,Collias-Baird-Borggreve-1994-P,Kumar-1993-CP}.
Therefore, microcellular polymer has wide industrial and
everyday-life applications including food packaging, airplane and
automotive parts, sporting equipment, insulation, controlled release
devices and filters, and so on
\cite{Suh-Park-Maurer-Tusim-Genova-Daniel-2000-AMater,Krause-Sijbesma-Munuklu-vanderVegt-Wessling-2001-Mm}.
The sizes and morphology of cells have considerable influence on the
application of foaming materials, which are usually studied using
the fracture surfaces
\cite{Park-1998-PES,Chandra-Gong-Turng-2004-JCP,Krause-Diekman-vanderVegt-Wessling-2002-Mm,Zhai-Yu-Wu-Ma-He-2006-P,Kraynik-Reinelt-Swol-2004-PRL,Hilgenfeldt-Kraynik-Reinelt-Sullivan-2004-EPL}.

In most cases, fracture surfaces of different materials are
self-similar, which can be characterized by fractal and multifractal
theories \cite{Mandelbrot-1983}. The monofractal properties of the
morphology of fracture surfaces have been investigated for metals
\cite{Mandelbrot-Passoja-Paullay-1984-Nature,Wendt-SL-Smid-2002-JM},
ceramics
\cite{Mecholsky-Passoja-FR-1989-JACS,Thompson-Anusavice-Balasubramaniam-Mecholsky-1995-JACS,Celli-Tucci-Esposito-Palmonari-2003-JECS},
polymers
\cite{Chen-Runt-1989-PC,Yu-Xu-Tian-Chen-Luo-2002-MD,Lapique-Meakin-Feder-Jossang-2002-JAPS,Zhou-Li-Liu-Cao-Zhao-Yuan-2006-PRE},
concretes
\cite{Dougan-Addison-2001-CCR,Wang-Diamond-2001-CCR,Issa-Issa-Islam-Chudnovsky-2003-EFM,Yan-Wu-Zhang-Yao-2003-CCC},
alloys
\cite{Bouchaud-Lapasset-Planes-1990-EPL,Shek-Lin-Lee-Lai-1998-JNCS,Wang-Zhou-Wang-Tian-Liu-Kong-1999-MSEA,Betekhtin-Butenko-Gilyarov-Korsukov-Lukyanenko-Obidov-Khartsiev-2002-TPL,Paun-Bouchaud-2003-IJF,Eftekhari-2003-ASS},
rocks
\cite{Schmittbuhl-Schmitt-Scholz-1995-JGRB,Lopez-Schmittbuhl-1998-PRE,Xie-Sun-Ju-Feng-2001-IJSS,Babadagli-Develi-2003-TAFM,Zhou-Xie-2003-SRL},
and many other materials. Moreover, the multifractal features of
surfaces have also been studies. Raoufi {\em{et al.}} analyzed the
multifractal spectrum of ITO thin films prepared by electron beam
deposition method and the $f(\alpha)$ shapes of ITO thin films
remained left hooked after annealing at $200^{\circ}\mathrm{C}$ and
$300^{\circ}\mathrm{C}$ \cite{Raoufi-Fallah-Kiasatpour-2008-ASS}.
Moktadir {\em{et al.}} researched the multifractal properties of
Pyrex and silicon surfaces blasted with sharp particles, and found
that the long-range correlations were the origin of the multifractal
behaviour \cite{Moktadir-Kraft-Wensink-2008-PA}.

In this work, we investigate the multifractal properties of the
fracture surfaces of polypropylene (PP) and polyethylene (PE) blends
foamed with supercritical carbon dioxide. The two-dimensional
multifractal detrended fluctuation analysis (MF-DFA) is adopted,
which has the advantages of easy implementation, high precision, and
low computational time \cite{Gu-Zhou-2006-PRE}. The MF-DFA approach
has been applied to investigate the landscape of the Yardangs region
on Mars and the fracture surface of a foamed polyurethane sample
with supercritical carbon dioxide
 \cite{Gu-Zhou-2006-PRE}, the combustion
flames in four-burner impinging entrained-flow gasifier
\cite{Niu-Zhou-Yan-Guo-Liang-Wang-Yu-2008-CEJ} and Pollocks's drip
paintings
\cite{AlvarezRamirez-LbarraValdez-Rodriguez-Dagdug-2008-PA}.

\begin{table}[t!]
  \flushleft
  \begin{tabular}{|ll|}
  \hline
    \noindent{\large\textbf{Nomenclature}}&\\
    $s$                                  & scale of boxes \\
    $X(i,j)$                             & two-dimensional matrix \\
    $u_{v,w}(i,j)$                   & cumulative sum \\
    $\widetilde{u}_{v,w}(i,j)$       & fitting bivariate polynomial \\
    $\epsilon_{v,w}(i,j)$            & residual matrix \\
    $q$                                  & order of detrended fluctuation function\\
    $F_{q}(s)$                           & detrended fluctuation function \\
    $D_{f}$                          & fractal dimension \\
    $h(q)$                               & scaling exponent function \\
    $\tau(q)$                            & mass exponent function \\
    $\alpha(q)$                          & singularity strength function \\
    $\alpha_{\max}$                   & maximum singularity \\
    $\alpha_{\min}$                   & minimum singularity \\
    $\Delta \alpha$                      & width of multifractal spectrum \\
    $f(\alpha)$                          & multifractal singularity spectrum    \\
    $\Delta f$                   & difference, $\Delta f=f(\alpha_{\max})-f(\alpha_{\min})$ \\
    $T$                                  & foaming temperature\\
    $w$                                  & weight fraction of PE\\
    $P$                                  & probability measure\\
  \hline
  \end{tabular}
\end{table}

\section{Experimental}

\subsection{Materials}

The isotactic polypropylene (Y1600) we have used was purchased from
Shanghai Petrochemical Company, China. The crystallinity and melting
temperature of the isotactic polypropylene was 47\% and
$169^{\circ}\mathrm{C}$, respectively.

The low density polyethylene (2426H) was purchased from Yangzi
Petrochemical Company, China. The crystallinity and melting
temperature of the low density polyethylene was 42\% and
$111.2^{\circ}\mathrm{C}$, respectively.

The CO$_{2}$ (purity: 99.9\%) supplied by Shanghai Air Product
Company, China, was utilized as a blowing agent.

\subsection{Experiment process}

Five different weight fractions of the PP/PE blends were studied, in
which the PP fractions are 100\%, 95\%, 90\%, 75\% and 50\%. The
blends were prepared in a Haake Minilab system, which is based on a
conical twin-screw compounder with an integrated backflow channel.
The blending was carried out for 10 minutes under a 0.6 MPa nitrogen
atmosphere with a blending temperature of $190^{\circ}\mathrm{C}$
and a screw speed of 50 rounds per minute. After blending, the
rod-like PP/PE blends were collected at the die exit for foaming.

\begin{figure}[htb]
\centering
  \includegraphics[width=8cm]{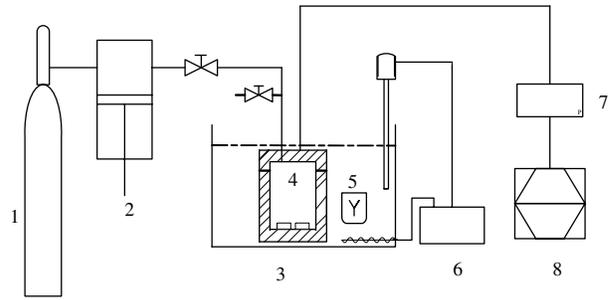}\\
  \caption{Schematic diagram of the experiment setup for the batch foaming process:
   1, CO$_{2}$ steel cylinder; 2, syringe pump; 3, oil bath; 4, high-pressure vessel;
  5, stirrer; 6, temperature controller; 7, pressure transducer; 8, data acquisition system.}
  \label{Fig:experiment}
\end{figure}

Fig.~\ref{Fig:experiment} illustrates the schematic experimental
setup of the depressurization batch foaming process. A high-pressure
vessel in stainless steel was used. The internal volume of the
vessel was 80 cm$^{3}$, calibrated with distilled water by a syringe
pump. A pressure transducer of type P31 from Beijing Endress \&
Hauser Ripenss Instrumentation Company Limited, was used to measure
the pressure with a precision of $\pm 0.01$ MPa and a valve of type
Swagelok SS-1RS8MM to release the CO$_{2}$ gas. A computer installed
with a PCI bus data acquisition system was connected to the above
pressure transducer to record the pressure decay during a
depressurization process.

In the foaming process, all PP/PE blends samples (including the pure
PP sample for comparison) were placed in the high-pressure vessel to
ensure the same foaming condition and the vessel was purged with
low-pressure CO$_{\rm 2}$. Thereafter, a given amount of CO$_{2}$
was charged. The CO$_{2}$ loading was achieved by a DZB-1A syringe
pump of Beijing Satellite Instrument Company (China) with a
precision of 0.01 cm$^{3}$. The high-pressure vessel was immersed in
a silicone oil bath and rapidly heated to a desired saturation
temperature. After the sorption of CO$_{2}$ into the blends samples
reached a sort of equilibrium, the CO$_{2}$ was released rapidly
from the high-pressure vessel. The foamed blends samples were taken
out for subsequent analysis. The samples were immersed in liquid
nitrogen for 10 minutes and then fractured. The cell morphologies of
the foamed blends samples were characterized by a JSM-6360LV
scanning electron microscopy (SEM). The polymer foaming was
influenced by many factors. In this work, we consider the influence
of different mixture ratios at two different temperature $T=150$ and
$140^{\circ}\mathrm{C}$.

\subsection{SEM images}

\begin{figure}[htb]
\centering
  \includegraphics[width=7cm]{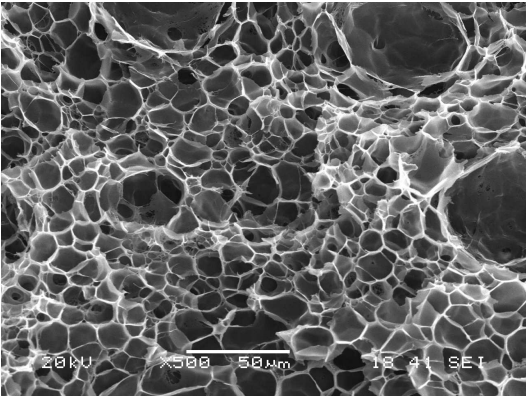}
  \caption{SEM micrograph of the foamed sample (${\rm{PP}}:{\rm{PE}}=90:10$) under condition:
  $150^{\circ}\mathrm{C}$, 25 MPa, depressurization rate 200 MPa/s. }\label{Fig:pic7-1}
\end{figure}

From the experiments, many SEM pictures of the fracture surface of
the foams under different experimental conditions were obtained.
Fig.~\ref{Fig:pic7-1} illustrates a typical image of fracture
surface of the foamed sample prepared at $150^{\circ}\mathrm{C}$ and
25 MPa with a depressurization rate of 200 MPa/s and the mixture
ratio ${\rm{PP}}:{\rm{PE}}=90:10$. The depressurization rate was not
a constant during the depressurization process. The rate of 200
MPa/s was the largest depressurization rate during the whole
depressurization process, which is the key parameter mainly
determining the cell nucleation rate. The data acquisition system
obtained the change of pressure with time and then we were able to
calculate the depressurization rate at any time, such that the
depressurization rate can be calculated as the ratio of the pressure
difference over the depressurization time. The size of the images
was $960\times 1280$ pixels, and we intercepted them into $800\times
1200$ to eliminate the noise of the mark on the images. The images
were stored in the computer as two-dimensional arrays in 256 grey
levels for multifractal analysis.

\begin{figure}[htb]
\centering
  \includegraphics[width=7cm]{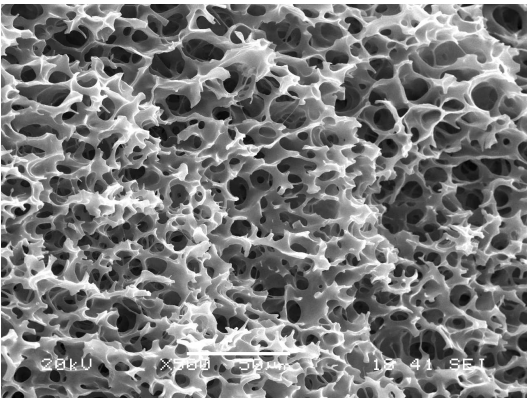}
  \caption{SEM micrograph of the foamed sample (${\rm{PP}}:{\rm{PE}}=50:50$) under condition:
  $150^{\circ}\mathrm{C}$, 25 MPa, depressurization rate 200 MPa/s.}\label{Fig:pic9-3}
\end{figure}

Fig.~\ref{Fig:pic9-3} illustrates a typical fracture surface image
of the foamed sample prepared at $150^{\circ}\mathrm{C}$, 25 MPa,
with the depressurization rate of 200 MPa/s, and the mixture ratio
${\rm{PP}}:{\rm{PE}}=50:50$. And we can find the open-cell easily,
and the same result can be obtained at $140^{\circ}\mathrm{C}$.
Thus, we can imagine that there is a threshold content of PE in the
sample at the range $25\%\sim50\%$ where the cells change to open.

\section{Two-dimensional multifractal detrended fluctuation analysis}

The multifractal properties may exist in the fracture surface of the
foams. To unveil the multifractal characteristics of the SEM images,
the two-dimensional MF-DFA is applied. This method has become
popular for its simplicity and easy computer implementation. The
two-dimensional MF-DFA is the extension of the DFA method which is
frequently used in computing the roughness exponent of monofractal
signals and the identification of long range correlations in
non-stationary time series
\cite{Kantelhardt-Zschiegner-Bunde-Havlin-Bunde-Stanley-2002-PA}.
The two-dimensional MF-DFA applied to our experiment data can be
summarized as following steps:

{\bf{Step 1.}} Consider a self-similar (or self-affine) surface,
which is denoted by a two-dimensional array $X(i,j)$, where $i=1,2,
. . . ,M$ and $j=1,2, . . . ,N$. The surface is partitioned into
$M_s\times N_s$ disjoint square segments of the same size $s\times
s$, where $M_s=[M/s]$ and $N_{s}=[N/s]$. Each segment can be denoted
by $X_{v,w}$ such that $X_{v,w}(i,j)=X(l_{1}+i,l_{2}+ j)$ for
$1\leqslant i,j \leqslant s$, where $l_{1}=(v-1)s$ and
$l_{2}=(w-1)s$.

Since $M$ and $N$ are often not a multiple of the segment size $s$,
two orthogonal strips at the end of the profile may remain and some
data will be ignored by this way. In order to take these ending
parts of the surface into consideration, the same partitioning
procedure can be repeated starting from the other three corners.
Then there will be $4M_sN_s$ segments and calculating the average
over them can eliminate the boundary influence.

{\bf{Step 2.}} For each segment $X_{v,w}$ identified by $v$ and $w$,
the cumulative sum $u_{v,w}(i,j)$ is calculated as follows:
\begin{equation}\label{1}
    u_{v,w}(i,j)=\sum\limits_{k_{1}=1}^{i} \sum\limits_{k_{2}=1}^{j}
    X_{v,w}(k_{1},k_{2})~,
    \end{equation}
where $1\leqslant i,j\leqslant s$. Note that $u_{v,w}$ itself is a
surface.

{\bf{Step 3.}} The trend of the constructed surface $u_{v,w}$ can be
determined by fitting it with a prechosen bivariate polynomial
function $\tilde{u}_{v,w}(i,j)$. The parameters of
$\tilde{u}_{v,w}(i,j)$ can be estimated easily through the least
square method fit to the data in each segment $X_{v,w}$. Then, the
residual matrix can be obtained. The detrended fluctuation function
$F(v,w,s)$ of the segment $X_{v,w}$ is defined via the sample
variance of the residual matrix:
\begin{equation}\label{2}
    F^2(v,w,s)=\frac{1}{s^{2}}\sum\limits_{i=1}^{s}\sum\limits_{j=1}^{s}
    \left[u_{v,w}(i,j)-\tilde{u}_{v,w}(i,j)\right]^2~.
\end{equation}

{\bf{Step 4.}} The overall detrended fluctuation is calculated by
averaging over all the segments,and the value of the $q$th-order
fluctuation function is
\begin{equation}\label{3}
    F_{q}(s)=\left\{\frac{1}{M_sN_s}\sum\limits_{v=1}^{M_{s}}\sum\limits_{w=1}^{N_s}[F(v,w,s)]^{q}\right\}^{1/q}~,
\end{equation}
where $q$ can take any real value except for $q=0$. When $q=0$,  we
have
\begin{equation}\label{4}
    F_{0}(s)=\exp\left\{\frac{1}{M_sN_s}\sum\limits_{v=1}^{M_{s}}\sum\limits_{w=1}^{N_{s}}\ln[F(v,w,s)]\right\}
\end{equation}
according to L'H\^{o}pital's Rule.

{\bf{Step 5.}} Varying the value of s in the range from $s_{\rm
min}\approx6$ to $s_{\max}\approx$ min($M,N$)/4, we can determine
the scaling relation between the detrended fluctuation function
$F_q(s)$ and the size scale $s$, which reads:
\begin{equation}
    \label{Eq:Fq:s:hq}
    F_q(s)\sim s^{h(q)}~,
\end{equation}
where the exponent $h(q)$ is called the generalized Hurst index. The
scaling exponent $h(q)$ is a constant for monofractals and a
nonlinear decreasing function of $q$ for multifractals. For positive
$q$ values, $h(q)$ describes the scaling behaviour of the segments
with large fluctuations, whereas for negative $q$ values $h(q)$
concerns with small fluctuations.

In the standard multifractal formalism based on partition function
\cite{Halsey-Jensen-Kadanoff-Procaccia-Shraiman-1986-PRA}, the
multifractal nature is characterized by the mass exponents
$\tau(q)$, which is a nonlinear function of $q$. The $\tau(q)$
function is related to $h(q)$ through
\begin{equation}
  \label{Eq:tau}
  \tau(q)=qh(q)-D_f
\end{equation}
where $D_f$ is the fractal dimension of the geometric support of the
multifractal measure
\cite{Kantelhardt-Zschiegner-Bunde-Havlin-Bunde-Stanley-2002-PA,Gu-Zhou-2006-PRE}.
According to the Legendre transform, we can obtain the singularity
strength function $\alpha(q)$ and the singularity spectrum
$f(\alpha)$ as follows
\cite{Halsey-Jensen-Kadanoff-Procaccia-Shraiman-1986-PRA}:
\begin{equation}\label{Eq:alpha}
    \alpha=\tau'(q)=h(q)+qh'(q)~,
    \end{equation}
\begin{equation}\label{Eq:f}
    f(\alpha)=q\alpha(q)-\tau(q)~.
\end{equation}

\section{Results}

\subsection{Multifractal analysis}

\begin{figure}[htb]
\centering
  \includegraphics[width=7cm]{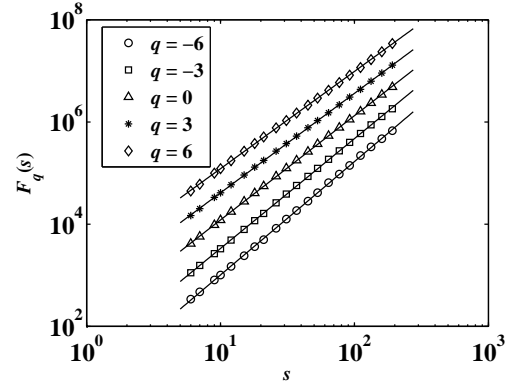}
    \caption{Log-log plots of $F_q(s)$ versus $s$ for five different values of $q$.
    The lines are the best power-law fits to the data. The plots for $q=-3$, 0, 3 and $6$ are
    shifted upward by $0.4, 0.8, 1.2$ and $1.6$ for clarity.}
    \label{Fig:Fq:s}
\end{figure}

As a first step, we perform multifractal analysis on the SEM images
to check if the fracture surfaces possess multifractal nature or
not. We take the SEM picture showed in Fig.~\ref{Fig:pic7-1} as an
example to illustrate the power-law scaling between the detrended
fluctuation function $F_q(s)$ and the scale $s$. Fig.~\ref{Fig:Fq:s}
shows the fluctuation $F_q(s)$ as a function of $s$ for five
different values of $q$ in double-logarithmic coordinates. The
results for $q=-3$, 0, $3$ and $6$ have been shifted upward by
$0.4$, $0.8$, $1.2$ and $1.6$ for clarity. According to the figure,
the data points for every $q$ fall on a straight line, indicating a
perfect power-law scaling between $F_q(s)$ and $s$, as expressed in
Eq.~(\ref{Eq:Fq:s:hq}). The scaling range is from 6 pixels to 200
pixels for all $q$ values. The fluctuation functions $F_q(s)$ of
other fracture surfaces also exhibit nice power-law behaviors.

\begin{figure}[htb]
\centering
  \includegraphics[width=8cm]{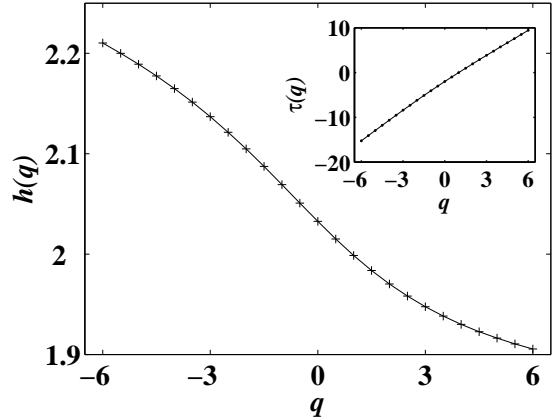}
  \caption{Dependence of the scaling exponent $h(q)$ with respect to the order $q$.
  The inset shows the mass exponent function $\tau(q)$.}
  \label{Fig:hq}
\end{figure}

According to Eq.~(\ref{Eq:Fq:s:hq}), the slopes of the straight
lines illustrated in Fig.~\ref{Fig:Fq:s} are the scaling exponent
$h(q)$, which can be determined by simple linear regressions of $\ln
F_q(s)$ against $\ln s$ for different $q$. Fig.~\ref{Fig:hq}
illustrates $h(q)$ as a function of $q$ for $-6\leqslant
q\leqslant6$ with the same image. It is noteworthy to point out
that, due to the finite size of the data, the overall fluctuation
values $F_q(s)$ for too large $q$ are not statistically significant.
It is evident that $h(q)$ is a decreasing nonlinear function of $q$.
The mass exponent function $\tau(q)$ is also calculated numerically
using Eq.~(\ref{Eq:tau}), where $D_f=2$. The inset of
Fig.~\ref{Fig:hq} shows the mass exponent function $\tau(q)$ for
$-6\leqslant q\leqslant6$. The nonlinearity of $h(q)$ and $\tau(q)$
confirms that the fracture surface under investigation possesses
multifractal nature. We note that all other fracture surfaces also
exhibit multifractality.

\subsection{Multifractal spectra}

Theoretically, the geometric support of a multifractal measure can
be decomposed into interwoven fractal sets, each of which is
characterized by its singularity strength $\alpha$. The fractal
dimension of the underlying fractal set associated with $\alpha$ is
$f(\alpha)$, which is the well-known singularity spectrum or
multifractal spectrum. In this vein, $\alpha$ and $f(\alpha)$ are
two most important characteristics in the description of the
multifractal.

\begin{figure}[htb]
\centering
  \includegraphics[width=8cm]{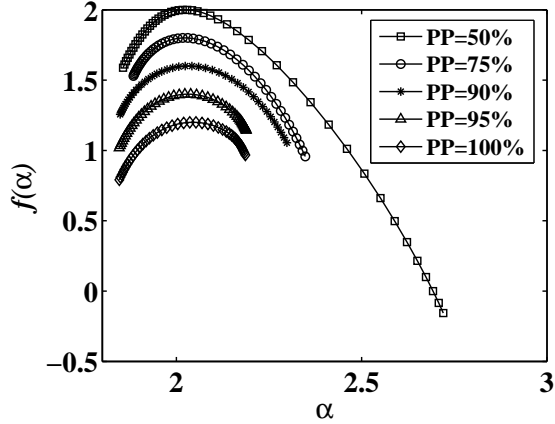}
 \caption{Multifractal spectra of the fracture surfaces of different PP/PE
 samples prepared at $T=150^{\circ}\mathrm{C}$.
 The curves for ${\rm{PP}}=75\%$, 90\%, 95\%, and 100\% have been
 shifted downward by 0.2, 0.4, 0.6 and $0.8$ for clarity.}
 \label{Fig:MFspectra:150}
\end{figure}

We have numerically obtained the values of $\alpha(q)$ and
$f(\alpha)$ through the Legendre transform from $\tau(q)$ for all
the fracture surfaces under investigation.
Fig.~\ref{Fig:MFspectra:150} illustrates the multifractal spectra
$f(\alpha)$ with respect to $\alpha$ for five samples with different
ratios for PP and PE at $150^{\circ}\mathrm{C}$. The multifractal
spectra for different blends samples exhibit different shapes. With
the increase of the proportion of polyethylene, the singularity
spectrum becomes wider. It means that the surface with higher
polyethylene proportion is more irregular, as is clear from the
comparison between Fig.~\ref{Fig:pic7-1} and Fig.~\ref{Fig:pic9-3}.
Another intriguing feature in Fig.~\ref{Fig:MFspectra:150} is that
the $f(\alpha)$ function for ${\rm{PP}}=50\%$ becomes negative when
$\alpha$ is larger than about 2.6. The negative dimension
($f(\alpha)<0$) was investigated in several experiments such as the
diffusion-limited aggregation
\cite{Amitrano-Coniglio-diLiberto-1986-PRL} and the energy
dissipation field of turbulent flows
\cite{Chhabra-Sreenivasan-1991-PRA}. The negative dimension
describes the rarely occurring events \cite{Mandelbrot-1990a-PA} and
one needs an exponentially increasing number of samples to observe
the subsets with the same $\alpha$ value
\cite{Chhabra-Sreenivasan-1991-PRA}.

\begin{figure}[htb]
\centering
  \includegraphics[width=7cm]{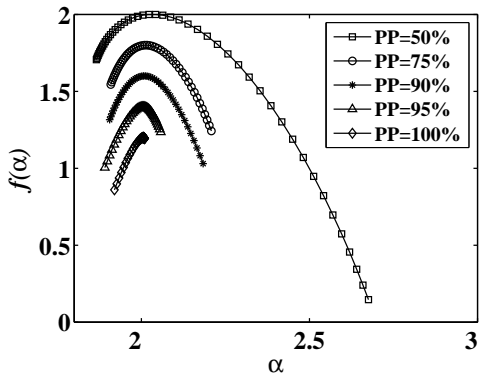}
 \caption{Multifractal spectra of the fracture surfaces of different PP/PE samples prepared at $T=140^{\circ}\mathrm{C}$.
 The curves for ${\rm{PP}}=75\%$, 90\%, 95\%, and 100\% have been shifted downward by 0.2, 0.4, 0.6 and $0.8$ for clarity.}
 \label{Fig:MFspectra:140}
\end{figure}

Fig.~\ref{Fig:MFspectra:140} plots the multifractal spectra
$f(\alpha)$ of the singularity $\alpha$ for five samples with
different PP/PE ratios at temperature $T=140^{\circ}\mathrm{C}$.
Again, the multifractal spectra for different blends samples exhibit
different shapes and the singularity spectrum becomes wider with the
increase of the proportion of polyethylene. However, no negative
dimension is observed in this case.

\section{Discussion}

There are several fundamental quantities related to the multifractal
spectrum. The minimum and maximum singularities $\alpha_{\min}$ and
$\alpha_{\max}$ are the singularity strengthes associated with the
regions of the sets where the measures are the least and most
singular, respectively. The corresponding $f(\alpha_{\min})$ and
$f(\alpha_{\max})$ values reflect the fractal dimensions of the two
regions characterized by $\alpha=\alpha_{\min}$ and
$\alpha=\alpha_{\max}$. The shape of the multifractal spectrum
$f(\alpha)$ can be captured to a great extent by the width of the
multifractal spectrum $\Delta\alpha=\alpha_{\max}-\alpha_{\min}$ and
the difference of the fractal dimensions
$\Delta{f}=f(\alpha_{\max})-f(\alpha_{\min})$. We discuss the
dependence of $\Delta\alpha$ and $\Delta{f}$ with respect to the
PP/PE ratio.

\subsection{The dependence of $\Delta\alpha$ with respect to the
PP/PE ratio}

In the formalism of multifractal, $\alpha_{\min}$ is related to the
maximum probability measure by
$P_{\max}\sim\varepsilon^{\alpha_{\min}}$, where $\varepsilon$
represents the scale approaching zero and it is a small quantity,
whereas $\alpha_{\max}$ is related to the minimum probability
measure through $P_{\min}\sim\varepsilon^{\alpha_{\max}}$. The width
$\Delta\alpha$ can be used to describe the range of the probability
measures:
\begin{equation}
 P_{\rm max}/P_{\rm min}\sim\varepsilon^{-\Delta\alpha}~.
 \label{Eq:dA:P}
\end{equation}
The greater the $\Delta\alpha$ value, the wider the probability
distribution and the larger the foamed growth probability of the
surface.

\begin{figure}[htb]
\centering
  \includegraphics[width=8cm]{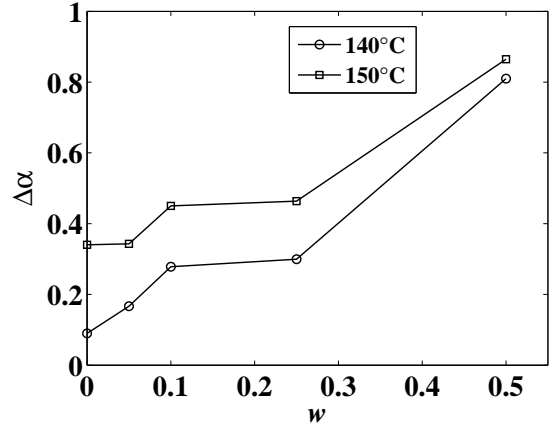}
  \caption{The width of spectrum $\Delta\alpha$ versus the content of PE ($w$) for two temperatures.}
  \label{Fig:deltaalpha}
\end{figure}

Fig.~\ref{Fig:deltaalpha} illustrates the relationship between
$\Delta\alpha$ and the content of PE in the sample at two different
temperatures. We find that the value of $\Delta\alpha$ increases
with augmenting PE accretion. This well agrees with the experimental
results that the number of cells in the sample increases with the
increase of the PE content, and the fracture surface becomes more
irregular. We also find that the two curves at different
temperaments share almost the same variation tendency. Since PP is a
semicrystalline polymer with a high crystallinity degree, it needs a
very rigorous condition for PP foaming since gases do not dissolve
in the crystalline regions
\cite{Naguib-Park-Reichelt-2004-JAPS,Naguib-Park-Song-2005-IECR,Patrick-Wang-Park-2006-IECR}.
According to Fig.~\ref{Fig:deltaalpha}, the $\Delta\alpha$ value of
the pure PP sample is merely 0.089, which means that the fracture
surfaces of pure PP samples are close to monofractal. When PE is
added in the sample, the PE with a low crystallization
temperature(Tc) can be molten (the melting point of LDPE we used is
$111.2^{\circ}\mathrm{C}$) in $140^{\circ}\mathrm{C}$, and both the
cell nucleation and bubble growth may take place in the region of
the molten PE. Therefore, mixing with PE makes the sample foaming
easier, and we obtain wider multifractal spectra with the increase
of the PE content while the crystalline regions may remain intact.
Comparing the two curves in Fig.~\ref{Fig:deltaalpha}, the curve for
foaming at $T=150^{\circ}\mathrm{C}$ is above that of
$T=140^{\circ}\mathrm{C}$, which is consistent with the results that
higher temperatures cause larger foamed regions with more cells and
larger cell size \cite{Xu-Jiang-Liu-Hu-Zhao-Zhu-Yuan-2007-JSF}.

According to Fig.~\ref{Fig:deltaalpha}, $\Delta\alpha$ increases
very quickly when the content of PE is more than 25\%. The cell
opening may be the reason for this transilient phenomenon. In our
experiment, the soft sections (PE) form minor and dispersed phases
and the hard sections (PP) form a major melt matrix. Cell opening
can be initiated and propagated through well-dispersed soft domains
that are entrapped between growing adjacent cells. Even though these
soft domains can become elongated as cells grow (i.e., cell walls
become thinner), cell opening is most likely to be initiated at the
weakest cell wall sections because of the embedded soft polymer
phases \cite{Patrick-Wang-Park-2006-IECR}. When the content of PE
reaches a certain value, the cell may be opening and continuous to
form a three-dimensional network, which leads to a much wider
distribution of the singularity of the fracture surface. We can
observe easily the open cells in Fig.~\ref{Fig:pic9-3} at
$T=150^{\circ}\mathrm{C}$, and at $T=140^{\circ}\mathrm{C}$ as well.

\subsection{The dependence of $\Delta{f}$ with respect to the
PP/PE ratio}

The parameter $\Delta f$ is also a very important quantity in the
multifractal analysis. The $f(\alpha_{\max})$ value reflects the
fractal dimension of the subset of the minimum growth probability
with
$N_{P_{\min}}=N_{\alpha_{\max}}\sim\varepsilon^{-f(\alpha_{\max})}$,
while $f(\alpha_{\min})$ reflects that of the maximum probability
such that
$N_{P_{\max}}=N_{\alpha_{\min}}\sim\varepsilon^{-f(\alpha_{\min})}$.
Hence, the $\Delta f(\alpha)$ value can describe the ratio between
the regions that the probability measure distributes most
concentrated and most rarified
\begin{equation}
 N_{P_{\max}}/N_{P_{\min}} \sim\varepsilon^{\Delta{f}}~.
 \label{Eq:df:P}
\end{equation}
Thus, $\Delta f <0$ means that there are more concentrated regions
than rarified sites, whereas $\Delta f >0$ means the contrary.

\begin{figure}[htb]
\centering
  \includegraphics[width=8cm]{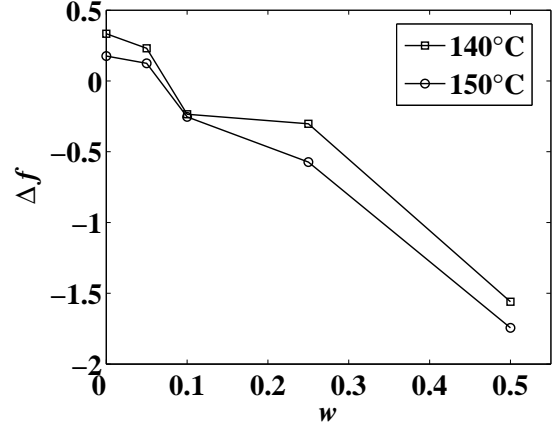}
  \caption{The difference of fractal dimension $\Delta f $ versus the content of PE ($w$)
  for two temperatures.}\label{Fig:deltaf}
\end{figure}

Fig.~\ref{Fig:deltaf} depicts the dependence of $\Delta{f}$ with
respect to the content of PE in the sample at two different
temperatures. With the increase of the PE content, the $\Delta f$
value changes from positive to negative, which reflects the change
of multifractal spectrum shape from left-hooked to right-hooked as
showed in Fig.~\ref{Fig:MFspectra:150} and
Fig.~\ref{Fig:MFspectra:140}. It implies that the fracture surface
becomes more complex and singular when there are more PE in the
sample.

The observed multifractal behaviour can be partly interpreted in
terms of the spatial intermittency
\cite{Argoul-Arneodo-Grasseau-1989-Nature,Kuramoto-Nakao-1997-PRL}
and the origin of the multifractal hidden in the fracture surface
may be the long range correlations of the intermittent fluctuation
\cite{Moktadir-Kraft-Wensink-2008-PA,Mordant-Delour-Leveque-Arneodo-Pinton-2002-PRL}.
A basic foaming process can be divided into three steps
\cite{Tomasko-Li-Liu-Han-Wingert-Lee-2003-IECR}: (1) mixing,
formation of a homogeneous solution composed of foaming agent and
polymer melt; (2) cell nucleation, phase separation induced by a
thermodynamic instability which is usually a temperature increase or
a pressure decrease; (3) cell growth and coalescence, a combination
of mass transfer and fluid dynamics. In bubble growth and
coalescence, the gas (CO$_{2}$) transferred from small bubble (with
high pressure) to big bubble (with low pressure) to form the foamed
structure
\cite{Leung-Park-Xu-Li-Fenton-2006-IECR,Leung-Li-Park-2007-JAPS},
which may be the origin of long-range correlation. However, in-depth
mechanism (mass transfer and fluid dynamics) research of foaming is
needed to support this surmise.

\section{Conclusion}

Foaming is a very complex process, and it is a challenging task to
underpin the exact mechanism explaining the nucleation and bubble
growth \cite{Leung-Park-Xu-Li-Fenton-2006-IECR}. In this work, we
obtained the foamed mixed polymers by batch foaming at fixed
pressure and depressurization rate. We used the two-dimensional
multifractal detrended fluctuation analysis to unveil the
multifractal properties of the fracture surfaces of formed PP/PE
blends samples. Perfect power-law scaling is observed and nonlinear
relationship between the scaling exponent $h(q)$ and $q$ in the
considering moment order $-6 \leqslant q \leqslant 6$ show that the
fracture surface exhibits multifractal nature.

The two important multifractal parameters $\Delta\alpha$ and $\Delta
f$ were calculated to describe the multifractal nature of the
fracture surfaces. We have found that, with the increase of the PE
fraction, the fracture surfaces become more and more irregular and
complex, which is indicated by the facts that $\Delta\alpha$
increases and $\Delta f$ decreases. These two parameters can serve
as ``complexity measures'' of the fracture surfaces of foamed
polymer blends, and other surfaces as well.

\bigskip
{\textbf{Acknowledgments:}}

We are grateful to Zhi-Qiang Jiang and Gao-Feng Gu for discussions.
This work was partly supported by the National Basic Research
Program of China (No. 2004CB217703), the National Natural Science
Foundation of China (Grant No. 50703011), the Program for Changjiang
Scholars and Innovative Research Team in University (IRT0620), the
Program for New Century Excellent Talents in University
(NCET-07-0288), the NSFC/PetroChina through a major joint project on
multiscale methodology (No. 20490200), and the Project Sponsored by
the Scientific Research Foundation for the Returned Overseas Chinese
Scholars, State Education Ministry.

\bibliography{E:/Papers/Auxiliary/Bibliography}

\end{document}